\documentclass[preprint,amsmath,amssymb]{revtex4-1}
\usepackage{graphicx}
\usepackage{dcolumn}
\usepackage{bm}
\usepackage{latexsym}
\usepackage{amsfonts}
\usepackage{mathrsfs}
\usepackage{amssymb}
\usepackage{amsmath}
\usepackage{verbatim}

\newcommand{\DM}{dark matter\,}
\newcommand{\half}{\frac{1}{2}}
\newcommand{\AddrFreiburg}{
 Physikalisches Institut, Albert-Ludwigs-Universit\"at Freiburg - Fakult\"at f\"ur Mathematik und Physik, D-79104 Freiburg, Germany
 \medskip
 }

\begin{document}

\title{The scalar Singlet-Triplet Dark Matter Model}

\author{O.~Fischer} 
\altaffiliation[e-mail: ]{oliver.fischer$@$physik.uni-freiburg.de}
\author{J.J.~van~der~Bij}
\altaffiliation[e-mail: ]{vdbij$@$.uni-freiburg.de}
\affiliation{\AddrFreiburg}


\begin{abstract}
We consider a model for cold dark matter, which combines a real scalar singlet and a real scalar $SU(2)_L$ triplet field, both of which are residing in the odd representation of a global $Z_2$ symmetry. The parameter space of the model is constrained by the inferred dark matter abundance from the WMAP and Planck data, the most recent results from the direct dark matter search experiment LUX, the Z boson decay width from LEP-I and perturbativity of the coupling parameters. The phenomenology of the remaining parameter space is studied. We find that the model allows for DM masses near the electroweak scale and a variety of decay scenarios.
\end{abstract}


\maketitle
\flushbottom

\section{Introduction}
One of the most astounding revelations of the twentieth century for our understanding of the Universe was the discovery of non-baryonic dark matter \cite{Zwicky:1933gu,Rubin:1970zza}, which is about five times more abundant than baryonic matter. Up to now, \DM is undetected in the laboratory. It also drives structure formation on large scales and determines galactic and extra-galactic dynamics. The interpretation of \DM as representing a type of elementary particle, is the most studied and most successful. Since the standard model (SM) of particle physics does not provide a viable candidate for a dark matter particle, an extension of the SM is necessary. In general a Weakly Interacting Massive Particle (WIMP) is the preferred object, with masses between about ten and a few hundred GeV.

Aside from its precise nature, which will be difficult to establish, a new unbroken symmetry seems to be fundamental for dark matter particles. If the latter are charged under this symmetry they can be stable on cosmological time scales. Most of the popular \DM models start with a motivation, which is not directly related to the \DM problem, but nevertheless leads to a natural inclusion of suitable particle candidates. 

Commonly  supersymmetry is invoked as a principle, where then R-parity is taken as the conserved symmetry. In this case the lightest supersymmetric particle would be the dark matter candidate and a relation between supersymmetry and dark matter is claimed to exist. This argument is misleading. R-parity is an independent principle, introduced to avoid supersymmetric rapid proton decay. Since supersymmetry by itself does not imply R-parity, it is not supersymmetry that gives a natural dark matter candidate.

If simplicity is invoked as the main principle, scalar singlet fields \cite{Silveira:1985rk,McDonald:1993ex,Binoth:1994pv,Binoth:1996au,Burgess:2000yq,Boehm:2003hm}
are preferred as the simplest candidate for dark matter.

The next simplest candidate for \DM is given by a real scalar $SU(2)_L$-triplet field, as presented and studied in Refs.~\cite{Barbieri:2006dq,FileviezPerez:2008bj,Cirelli:2005uq}. This particle is a suitable \DM candidate, because radiative corrections to its mass render the neutral component lighter than the charged ones. 

Besides the precise nature of the dark matter particle it is a priori unclear whether the dark matter consists of one or more components. An interesting suggestion for a multi-component form of dark matter was made in \cite{Kadastik:2009dj}. It was argued that the dark matter should be part of a matter-parity odd representation of the unification group SO(10), the smallest being the {\bf 16}, which led to a combination of an inert doublet and a complex singlet. In the original motivation for this model, results from ref.\cite{vanderBij:2007fe,vanderBij:2010nu} were used. In ref.\cite{vanderBij:2007fe,vanderBij:2010nu} an argument was presented, that the fermions should come in exactly three generations of a {\bf 16}-spinor representation of $SO(10)$~\cite{vanderBij:2007fe,vanderBij:2010nu}. 
However it was found that the unification group should be $SU(5)$ and not $SO(10)$. 
When a matter parity is included in such a unification scenario, it is very likely that a larger number of fields reside in  matter-parity odd representations, which implies a multicomponent dark matter scenario. In particular, when the {\bf 24}-representation of $SU(5)$ is considered, beyond the Standard Model fields, a real singlet plus a $SU(2)_L$ triplet field are found. Since no experiment has detected such fields up to now, it is straightforward to assume that they have odd matter-parity, and thus might yield viable dark matter candidates.  Both fields couple to the Higgs boson, so that they scatter off terrestrial nuclei, which makes them sensitive to the null-results coming from direct detection experiments~\cite{Aprile:2011hi}.

The combination of singlet and triplet \DM fields will be studied in the following. Contrary to previous work in Ref.~\cite{Fischer:2011zz}, we postulate {\it one} \DM symmetry, which leads to interesting phenomenological implications.

This paper is structured in the following way:
In the next section, we introduce the singlet-triplet $Z_2$ model as an extension of the SM, which includes one global $Z_2$ symmetry and a real scalar singlet and a  real scalar $SU(2)_L$ triplet field to the matter content of the Theory. Including all the renormalizable terms in the Lagrangian leads to mass mixing of the singlet and triplet fields after spontaneous symmetry breaking.
In section~\ref{sec_DM} we study the effect of the abundance constraint on the parameter space of the model by employing the numerical tool micrOMEGAs~\cite{Belanger:2001fz,Belanger:2004yn}. This leads to three distinct scenarios, namely the singlet, the  triplet and 
the mixed scenario, which will be discussed in separate subsections. Section~\ref{sec_Conclusion} contains our summary and the conclusions.

\section{The Model}
The matter content of the SM is extended with two real scalar fields: a singlet $\Phi$ and a $SU(2)_L$ triplet. Both fields reside in the odd representation of an additional, global $Z_2$ symmetry, while the SM fields reside in the even representation. 

The Lagrangian of the singlet-triplet $Z_2$ model reads:
\begin{equation}
\mathscr{L}_{Z_2} = \mathscr{L}_{\rm SM} + \mathscr{L}_{\Phi} + \mathscr{L}_{\Psi} + \mathscr{L}_{\rm mix}\,,
\end{equation}
where $\mathscr{L}_{\rm SM}$ contains the usual SM fields and
\begin{eqnarray}
\mathscr{L}_{\Phi} & = & \half \partial_\mu \Phi \, \partial^\mu \Phi - \half m_{\Phi}^2 \Phi^2 -\frac{\lambda_{\Phi}}{4!} \Phi^4 - \frac{\omega_{\Phi}}{4} \Phi^2 H^{\dagger}H \,,\\
\mathscr{L}_{\Psi} & = & \half \mathcal{D}_{\mu} \Psi^{\dagger} \mathcal{D}^{\mu} \Psi - \half m_{\Psi}^2 \left|\Psi\right|^2  - \frac{\lambda_{\Psi}}{24}\left|\Psi\right|^4 -\omega_{\Psi} \left| \Psi \right|^2 H^{\dagger}H\,,\\
\mathscr{L}_{\rm mix} & = &  \kappa H^{\dagger} \tau^i H \Psi_i \, \Phi + \kappa' \Phi^2 |\Psi|^2\,. \label{eq:z2lgrgn}
\end{eqnarray}
In the above equations, $H$ is the SM Higgs doublet, $\lambda_\Phi,\lambda_\Psi,$
$\omega_\Phi,\omega_\Psi$ and $\kappa$ are dimensionless coupling constants, $m_\Phi,m_\Psi$ are the Lagrange masses of the singlet and triplet respectively. In the following, the coupling parameters $\lambda_{\Phi},\lambda_\Psi,\kappa'$ will be neglected, since their contribution to the \DM abundance was shown to be irrelevant \cite{Fischer:2011zz}.

The spontaneous breaking of the electroweak symmetry leads to a mixing of the neutral flavour eigenstates $\Phi,\,\Psi^0$ through the first term in eq.~(\ref{eq:z2lgrgn}). The mixed fields $S_1$ and $S_2$ are linear combinations of the interaction eigenstates, which can be expressed as
\begin{equation}
\left(\begin{array}{c} \Phi \\ \Psi^0 \end{array} \right) = \left( \begin{array}{cc} c_{\delta} & s_{\delta} \\ -s_{\delta} & c_{\delta} \end{array} \right) \left(\begin{array}{c} S_1 \\ S_2 \end{array} \right),
\label{SDM_z2_rot}
\end{equation}
where $s_{\delta}\,(c_{\delta})$ are the sine (cosine) of the mixing angle $\delta$. For conformity, the charged triplet components are relabeled to $S^{\pm} \equiv \Psi^{\pm}$ and the mass $m_c \equiv m_\Psi+\Delta m$, where $\Delta m$ is the loop-induced mass splitting due to the gauge couplings of the triplet field~\cite{Cirelli:2005uq}.

\begin{figure}
\begin{center}
\includegraphics[width=0.4\textwidth,angle=-90]{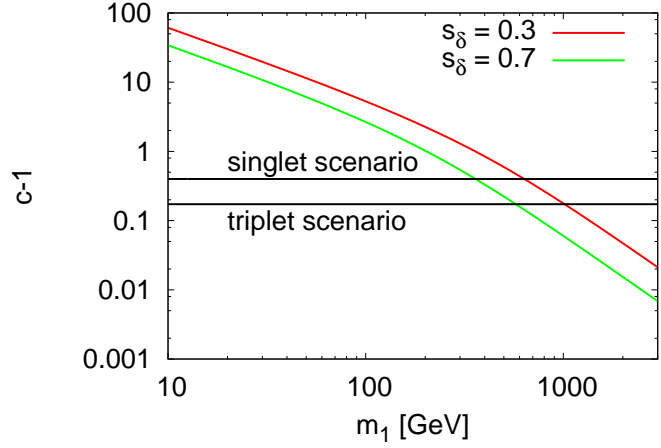}
\end{center}
\caption{The perturbative upper limit on the mass-splitting parameter $c$ for two different values of the singlet-triplet mixing angle $\delta$. The lines are upper exclusion contour lines.}
\label{fig-cmax}
\end{figure}

\begin{figure}
\begin{center}
\includegraphics[width=0.4\textwidth,angle=-90]{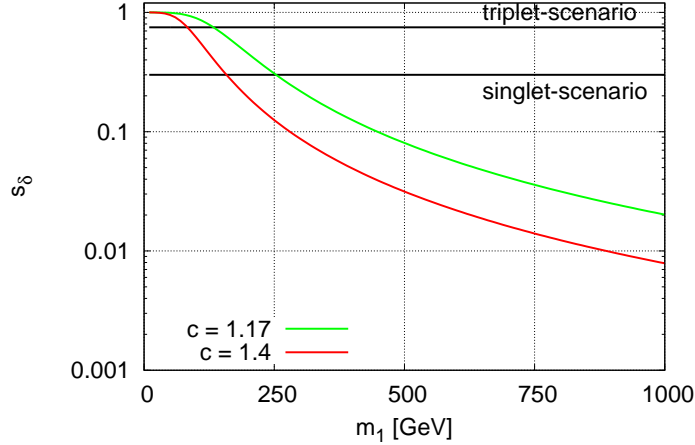}
\end{center}
\caption{The perturbative upper limit on the singlet-triplet mixing angle $\delta$ for two different values of the mass-splitting parameter $c$. The lines are upper exclusion contour lines.}
\label{fig-dmax}
\end{figure}

The physical masses of the fields $S_1,S_2,S^\pm$, or $S-$fields for brevity, are given by the eigenvalues of the mass matrix:
\begin{equation}
m_\pm^2 = \half \left[m_{\Phi}^2+m_{\Psi}^2 \pm \sqrt{\left(\Delta\right)^2  + \kappa^2v^4} \right]\,,  \label{SDM_z2_m1}
\end{equation}
with $\Delta = m_{\Phi}^2 - m_{\Psi}^2$. We identify $m_-,m_+$ with $m_1,m_2$ respectively. The mixing angle can be expressed in terms of the physical masses as
\begin{equation}
s_\delta^2 = \frac{1}{2}\left[1 - \sqrt{1 - \frac{\kappa^2\,v^4 }{(m_1^2-m_2^2)^2}}\right]\,.
\end{equation}
The mass splitting between $m_1$ and $m_c$ can be defined via a mass-splitting parameter $c$, as
\begin{equation}
c \equiv {m_c \over m_1}\,.
\label{SDM_z2_mc=cm1}
\end{equation}
We have the following useful relations between parameters:
\begin{eqnarray}
m_2^2 & = & m_1^2 \left[1 + \frac{c^2-1}{c_\delta^2} \right]   \label{SDM_z2_fm2} \\
\kappa & = & 2\,t_\delta \left(c^2-1\right)\frac{m_1^2}{v^2}\,. \label{SDM_z2_fkappa2}
\end{eqnarray}
Note that in the limit of $\delta$ going to zero, $m_2$ converges to $m_c$ and $\kappa$ to zero, as it should. It is convenient to define the mass splitting between $S^\pm$ and $S_2$, analogous to the definition in eq.~(\ref{SDM_z2_mc=cm1}), as
\begin{equation}
c_2 \equiv {m_2 \over m_1} = \frac{\sqrt{c^2 - s_\delta^2}}{c_\delta}\,.
\label{SDM_z2_c2}
\end{equation}
The couplings $\omega_{ij}$, defined via the terms $H S_i S_j$ are given by
\begin{eqnarray}
-\omega_{11} & \equiv & 4\,s_\delta c_\delta\kappa-c_\delta^2\omega_\Phi-s_\delta^2\omega_\Psi\,, \label{SDM_z2_OM1}\\
-\omega_{12} & \equiv & -2\,(s_\delta^2-c_\delta^2)\kappa-s_\delta c_\delta(\omega_\Phi-\omega_\Psi)\,, \\
-\omega_{22}  & \equiv & -4s_\delta c_\delta\kappa-s_\delta^2\omega_\Phi-c_\delta^2 \omega_\Psi\,.
\end{eqnarray}
While $\omega_\Phi,\omega_\Psi$ are free parameters, $\kappa$ is a function of the model parameters $m_1,c,\delta$. In defining the Lagrange parameters $\omega_\Phi,\omega_\Psi$ as functions of the effective Higgs couplings $\omega_{11},\omega_{22}$, the latter become the free parameters of the model. This choice of parameter set gives
\begin{eqnarray}
-\omega_{12} & = & \frac{2\kappa}{c_\delta^2-s_\delta^2} - \frac{s_\delta c_\delta \omega_{11}}{c_\delta^2-s_\delta^2} - s_\delta c_\delta \omega_{22}\,, \\
-\omega_\Psi & = & \frac{4 s_\delta c_\delta \kappa-c_\delta^2 \omega_{22} - s_\delta^2 \omega_{11}}{s_\delta^2 - c_\delta^2}\,.
\end{eqnarray}
The perturbativity of the theory is violated with $|\omega_{12}|,|\omega_\Psi| > 1$, which defines upper bounds for the model parameters $c,\delta$:
\begin{eqnarray}
c_{\rm max} & = & \sqrt{\frac{v^2}{t_\delta m_1^2}+1}\,, \label{SDM_z2_cmax} \\
\delta_{\rm max} & = & \arctan\left[\frac{v^2}{m_1^2\left(c^2-1\right)}\right]\,. \label{SDM_z2_deltamax}
\end{eqnarray}
Fig.~\ref{fig-cmax} shows the resulting bounds for the mass-splitting parameter $c$ as a function of the mass, for two different, allowed values of the mixing angle. For $s_\delta \to 1 (0)$, the upper bound becomes tighter (relaxed). Fig.~\ref{fig-dmax} shows the resulting upper bound on $s_\delta$ as a function of the mass, for two given values of the mass-splitting parameter $c$. An increased (reduced) value of $c$ implies a tighter (relaxed) upper bound on $s_\delta$ for a given mass. Note that the black lines represent the separation of the singlet and the triplet scenario as defined below. In Fig.~\ref{fig-dmax}, the excluded range of the mixing angle is within the black lines. 

Considering $Z$ boson decays into a pair of charged $S^\pm$, the associated partial decay width is ${\cal O}(1$ GeV), for $m_c$ not close to or above the decay threshold. The experimental limit is $\Gamma_{Z \to x\bar{x}} \leq 4.2$ MeV~\cite{Beringer:1900zz}, with $x$ being a non SM particle. This limit is a lower bound on the triplet mass:
\begin{equation}
m_c > \frac{m_Z}{2}\,.
\label{LEP-I}
\end{equation}

\section{Dark Matter Properties}
\label{sec_DM}
The \DM relic density can be inferred from WMAP~\cite{Hinshaw:2012aka} and Planck~\cite{Ade:2013zuv} data. An analysis which combines the results from both experiments yields the value
\begin{equation}
\Omega_{DM} h^2  =  0.1199 \pm 0.0027\,.
\label{eq-abundance}
\end{equation}
We shall use this value as constraint on the model parameters in the following, and quote it with {\it abundance} constraint.

The lightest $Z_2$ odd field $S_1$ is the \DM candidate of the singlet-triplet $Z_2$ model. It couples to the Higgs boson and the gauge fields. For a very small value of $\delta$ the Higgs coupling is dominant in the annihilation cross section of the $S_1$, so that the model is expected to behave similar to the pure scalar singlet model~\cite{Binoth:1994pv,Binoth:1996au,Silveira:1985rk,McDonald:1993ex,Burgess:2000yq,Boehm:2003hm}. For $\delta \to \pi/2$, the gauge couplings become dominant in the annihilation cross section of the $S_1$, making the model similar to the pure triplet model~\cite{Barbieri:2006dq,FileviezPerez:2008bj,Cirelli:2005uq}.

Considering the heavier $S_i,\,S_j$, their coannihilations are suppressed by a Boltzmann factor
\begin{equation}
\ln{B} = -X_f \, \left(\frac{m_i+m_j - 2\, m_1}{m_1}\right)\,,
\end{equation}
with the masses $m_i,\,m_j \in (m_2,\,m_c)$ and $X_f$ the freeze-out temperature of the $S_1$. The numerical tool MicrOMEGAs considers these processes up to $\ln{B} =-13.8$. For the coannihilations of the $S^\pm$, with $m_i=m_j=c\,m_1$, this yields the function for the mass-splitting parameter $c$:
\begin{equation}
f_c(X_f) = 1 + \frac{6.9}{X_f}\,.
\label{SDM_z2_ccrit}
\end{equation}
This function serves as a benchmark, whether $S^\pm$ coannihilations are being included in the computation of the abundance. Since typical WIMP freeze-out temperatures vary between 20 and 40, for values of the mass-splitting parameter given by $c \geq f_c(20) = 1.345$, coannihilations are not considered by micrOMEGAs. Conversely, for $c\leq f_c(40)= 1.1725$ the coannihilations of the other particles are included into the computation of the relic abundance.

\subsection{Singlet Scenario}
We define the {\it singlet scenario}, so that the $S_1$ abundance is dependent on the mass $m_1$ and the Higgs coupling $\omega_{11}$, comparable to the pure scalar singlet model as in Refs.~\cite{Silveira:1985rk,McDonald:1993ex,Burgess:2000yq}, in the following way:
\begin{equation}
s_\delta \leq 0.3 \qquad {\rm and} \qquad c \geq 1.4\,.
\end{equation}
The annihilation cross section is dominated by Higgs exchange, the strength of which is controlled by $\omega_{11}$. The coannihilations of the $S^\pm$ fields are negligible according to eq.~(\ref{SDM_z2_ccrit}). Notice that the choice for the limiting values for $s_\delta$ and $c$ is somewhat arbitrary. In this scenario, the abundance constraint can be matched by tuning the parameters $m_1,\,\omega_{11}$, and, to a lesser extent, $\delta$. 

\begin{figure}
\begin{center}
\includegraphics[width=0.4\textwidth,angle=-90]{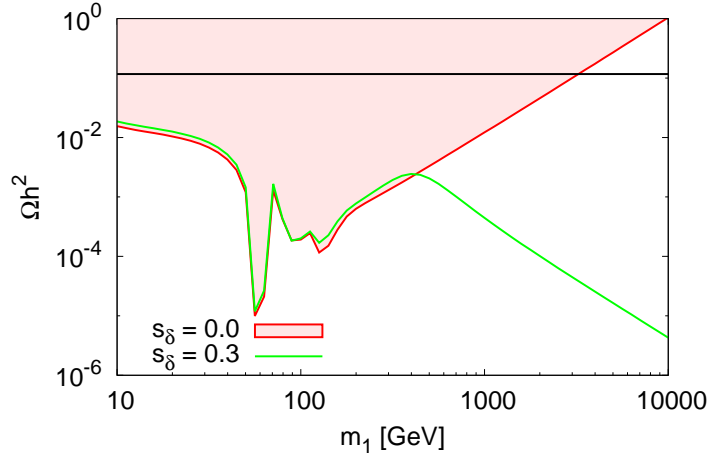}
\end{center}
\caption{Abundance of the singlet scenario as a function of the mass $m_1$. The red area represents values which are allowed by the constraints on the parameter space. For the green curve, $s_\delta=0.3$  has been used. For $m_1=457$ GeV the two lines diverge due to $\kappa > 1$, see eq.~(\ref{eq-m1bound}).}
\label{fig-singlet}
\end{figure}

The mass range for $m_1$ for a given $\delta$ is constrained by the perturbativity of the couplings:
\begin{equation}
m_1 < \frac{246 \text{ GeV}}{\sqrt{t_\delta(c^2-1)}}\,,
\label{eq-m1bound}
\end{equation}
which for $s_\delta = 0$ and $c\geq 1.4$ recovers the limits on the mass range from the pure singlet model, which comes from the constraint $|\omega_{11}|\leq 1$, and eq.~(\ref{eq-abundance}). This is illustrated by the red area in Fig.~\ref{fig-singlet}, which represents allowed values for the abundance. The mass is limited from above by $m_1 \leq 3.4$ TeV. The green curve in the figure represents the case of $s_\delta=0.3$ and $c=1.4$. It diverges from the red area at $m_1=457$ GeV, in agreement with the formal upper bound on $m_1$ from eq.~(\ref{eq-m1bound}). This divergence is due to $\omega_{12}$ becoming larger than one, growing proportional to $m_1^2$. This increases $\sigma_A$ through contributions from coannihilations of the form $S_1 S_2 \to h \to \ f\bar{f}$, suppressing the abundance. An increase of $c$ does increase the Boltzmann suppression, but at the same time it also increases the coupling $\omega_{12}$.

Considering the mass spectrum in the singlet scenario, the parameter $c\geq 1.4$ determines the mass splitting between the $S_1$ and the $S^\pm$ fields. The relative mass splitting between the $S^\pm$ and the $S_2$ is given by eq.~(\ref{SDM_z2_c2}). For $s_\delta=0$ it is $c_2=c$, and the two heavier fields are mass degenerate, apart from the radiative corrections to the $S^\pm$. The maximum value for $c_2$ is given with $c_2^{\rm max} = 1.05$ for large $c$ and $s_\delta=0.3$.

Subsequently, the $S_2,S^\pm$ form a set of fields that is similar to the three fields in the pure triplet model. We point out two differences. First, the $S_2$ tends to be heavier than the $S^\pm$. Second, the relative mass difference between $S_2$ and $S^\pm$ does not converge to zero for large masses.

\begin{figure}
\begin{center}
\includegraphics[width=0.4\textwidth,angle=-90]{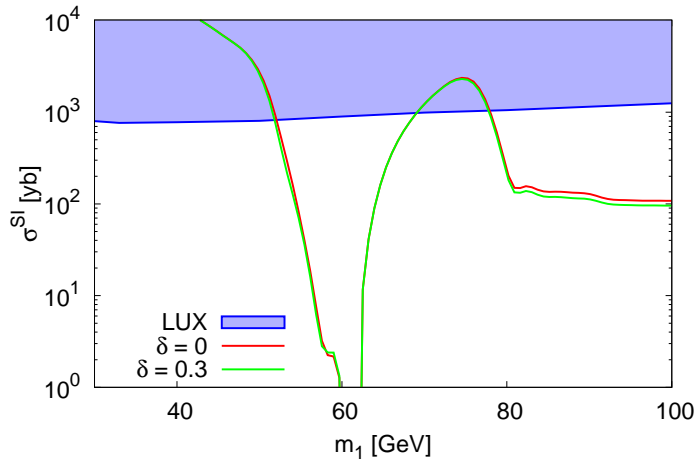}
\end{center}
\caption{Direct detection constraints on the mass parameter in the singlet scenario. The spin independent $S_1$-nucleon cross section is displayed in yoctobarn. The blue area represents the most recent exclusion limits from the LUX experiment~\cite{Akerib:2013tjd}, the red line represents the abundance constraint from WMAP and Planck data on the coupling of the $S_1$ to the Higgs boson.}
\label{fit-singletDS}
\end{figure}

In the singlet scenario, the coannihilations and the perturbative constraints on the parameter $c$ are negligible for $m_1 \sim m_H/2$. Therefore, the most recent constraints from the direct detection experiment LUX~\cite{Akerib:2013tjd} affect the mass range in the same manner as in the pure singlet model. This places the lower bound of $53$ GeV on $m_1$ and also excludes the mass range 66 GeV $\leq m_1 \leq$ 78 GeV, as shown in Fig.~\ref{fit-singletDS}. This range is not visibly affected by the variation of $s_\delta$ between zero and 0.3, so that the mass range for the $S_1$ is given by 78 GeV $\leq m_1 \leq$ 3.4 TeV, except for the narrow range around the Higgs resonance.

We note at this point that here and in the following, we use the value for the strangeness of the nucleon as reported in Ref.~\cite{Alarcon:2012nr}.

\subsection{Triplet Scenario}
\begin{figure}[h]
\begin{center}
\includegraphics[width=0.4\textwidth,angle=-90]{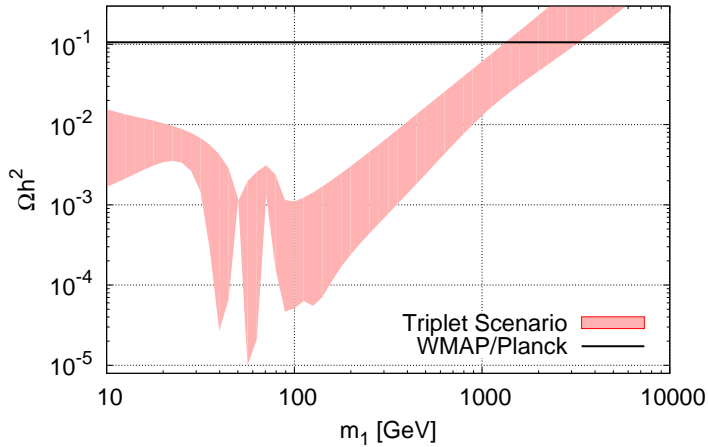}
\end{center}
\caption{Abundance of the triplet scenario as a function of the mass $m_1$. The red area represents values for the abundance, resulting from a maximum variation of the coupling constants $\omega_{11},\omega_{22}$, the singlet-triplet mixing angle $s_\delta$ and the mass-splitting parameter $c$. The black line represents the abundance constraint from WMAP~\cite{Hinshaw:2012aka} and Planck~\cite{Ade:2013zuv}, its uncertainty is too small to show in this plot.}
\label{SDM_z2_tripleteps}
\end{figure}
We define the {\it triplet scenario}, so that the abundance is qualitatively close to that of the pure triplet model as in Refs.~\cite{Barbieri:2006dq,FileviezPerez:2008bj,Cirelli:2005uq}, by the following constraints:
\begin{equation}
s_\delta \geq 0.75 \qquad {\rm and} \qquad c < 1.17\,.
\end{equation}
Because of the small value for $c$, the coannihilations of the $S^\pm$ fields are contributing to $\sigma_A$, resulting in an increased sensitivity of the abundance on the couplings $\omega_{22},\omega_{12},\omega_\Psi$, compared to the singlet scenario. The large value of $s_\delta$ yields strong gauge couplings of the $S_1$ field, which in turn reduces the sensitivity to $\omega_{11}$.

The masss-splitting parameter $c$ is very powerful in controlling the contributions from the $S^\pm,S_2$ to the abundance, and their coannihilation strength. For $c$ close to one, the Boltzmann suppression of the $S^\pm,S_2$ abundances becomes negligible, which contributes to, and thus increases the, $S_1$ abundance. For a suitable choice of $c,s_\delta$, the coannihilations of the $S_2$ still contribute to $\sigma_A$, while its abundance is still suppressed. In this extreme case, $\Omega$ can take smaller values compared to the abundance in the triplet model for the same mass. On the other hand, if $c\simeq1$, the abundance can be larger than in the triplet model.

This makes the abundance being mainly sensitive to the variation of $m_1$ and $c$, and only minorly to a tuning of the Higgs couplings. Fig.~\ref{SDM_z2_tripleteps} shows how the variation of the parameters $\omega_{11},\omega_{22},s_\delta,c$ affect the abundance. The upper boundary of the red area is given for $s_\delta = 0.75$ and $c,c_2 \simeq 1$. The lower boundary is given for $s_\delta =1-\varepsilon$, with $\varepsilon\to0$ for numerical stability, and $c \sim c_{\rm max}$. The variation of the model parameters allows for the $S_1$ mass in the range of 
\begin{equation}
1.33 \text{ TeV} \leq m_1 \leq 6.65 \text{ TeV}\,,
\label{SDM_z2_tripletrange}
\end{equation}
to match the abundance constraint. It is possible to constrain the mass range in eq.~(\ref{SDM_z2_tripletrange}) with gamma ray data from the HESS collaboration~\cite{Abramowski:2011hc}. Adopting an NFW profile, this translates into a limit on the mixing angle:
\begin{equation}
s_\delta < \left(\frac{\Omega_1}{\Omega_{\rm HESS}}\right)^{1 \over 4}\,.
\end{equation}
For $m_1$ between $\sim 2$ and $\sim 3$ TeV, the Hess constraints limit the angle from above with $s_\delta < 0.75$ and therefore exclude the triplet scenario. We emphasize however, that this constraint is strongly dependent on the choice of the halo profile model~\cite{Fan:2013faa,Cohen:2013ama}.

\begin{figure}[h]
\begin{center}
\includegraphics[width=0.4\textwidth,angle=-90]{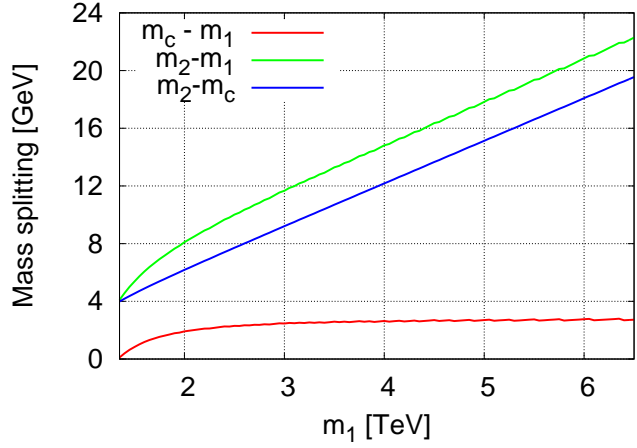}
\end{center}
\caption{The mass splittings $\Delta_{ij}$ for $i=2,c$ and $j=c,1$ between $S_1,S_2,S^\pm$ in the triplet scenario. The lines represent upper limits on the absolute values of the three different mass splittings.}
\label{SDM_z2_tripletdmax}
\end{figure}

Concerning the mass splitting between the three fields, we first consider the effect of the constraints on the parameter $c$. The abundance constraint limits the mass splitting parameter to $c \leq 1.001$. The upper limit for $m_1$ is given by (\ref{SDM_z2_tripletrange}). We define the absolute value of the mass splitting between the three masses with  
\begin{equation}
\Delta_{ij} = m_i - m_j\,.
\end{equation}
The upper limits for the $\Delta_{ij}$, with $i=2,c$ and $j=c,1$, are shown in Fig~\ref{SDM_z2_tripletdmax}. The absolute values of the upper bounds for the three parameters grow as large as $\sim 20$ GeV for $m_1 \to 6.5$ TeV. The ranges for the mass splittings are given by:
\begin{eqnarray}
0 & \leq \Delta_{21} & \leq 19.4 \text{ GeV}\,, \\
-0.168 \text{ GeV} &\leq \Delta_{2c} &\leq 16.6 \text{ GeV}\,, \\
0.168 \text{ GeV} &\leq \Delta_{c1} &\leq 2.8 \text{ GeV}\,,
\end{eqnarray}
which may have consequences for collider searches, as it opens up more decay channels.

\subsection{Intermediate Scenarios}
\begin{figure}[h]
\begin{center}
\includegraphics[width=0.4\textwidth,angle=-90]{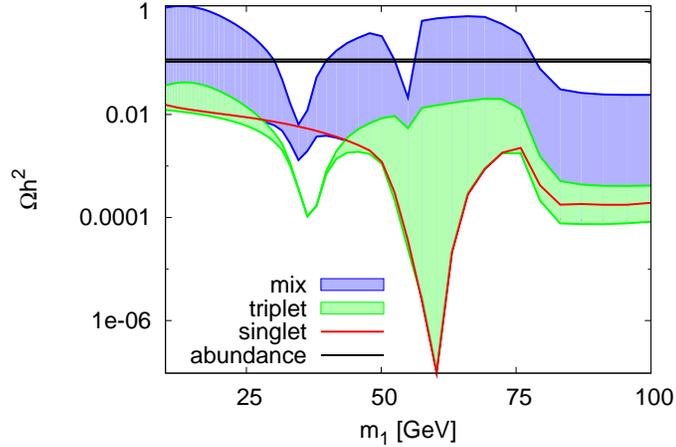}
\end{center}
\caption{The relic density as a function of mass $m_1$ for three different scenarios. The width of the blue and green area represents the variation of $\omega_{11},\omega_{22}$ between 0 and 1. For the singlet scenario $c=1.4$ and $s_\delta=0$ have been used, the intermediate scenario uses $c=1.2$ and $0.35 \leq s_\delta \leq 0.75$, excluding the range leading to $|\kappa|>1$. For the triplet scenario, $s_\delta = 0.9$ and $c=1.1$ were used.}
\label{SDM_z2_mixeps}
\end{figure}
\noindent
We define the {\it intermediate scenario} by the range that remains uncovered by the singlet and triplet scenario:
\begin{equation}
0.3 < s_\delta < 0.75 \qquad {\rm and} \qquad 1.17 < c < 1.4\,.
\end{equation}

The intermediate scenario is displayed in Fig.~\ref{SDM_z2_mixeps}, together with the singlet and the triplet scenario. We choose the mass range between 10 and 100 GeV to illustrate the effects of coannihilations and the gauge interactions of the $S_1$. The blue and green areas represent the intermediate and triplet scenario, respectively. The width of each area is given by the variation of the Higgs couplings $\omega_{11},\omega_{22}$, between 0 and 1. The red line represents the lower contour line for the singlet scenario. The black line denotes the abundance constraint from eq.~(\ref{eq-abundance}), which implies that the triplet scenario is excluded in this low mass range, while the intermediate scenario is able to match the constraint for a tightly constrained mass range.

\begin{figure}[h]
\begin{center}
\includegraphics[width=0.4\textwidth,angle=-90]{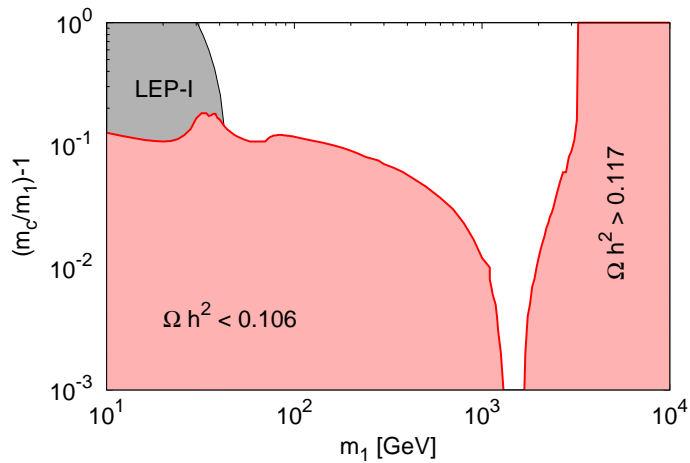}
\end{center}
\caption{Allowed range for the mass-splitting parameter $c$ as a function of the mass $m_1$. The red area represents parameter sets being excluded by WMAP and Planck data, the dark gray area represents the partial $Z$ boson decay width into $S^\pm$ violating LEP-I limits \cite{Beringer:1900zz}.}
\label{SDM_z2_m1ceps}
\end{figure}

The $Z$ and $W^\pm$ boson resonances are Boltzmann suppressed in the singlet scenario, their magnitude is proportional to $s_\delta$. The resonance of the weak gauge bosons is shifted to smaller values of $m_1$ due to the center-of-mass energy of the associated process, $S_1 S^\pm \to W^\pm \to f^0 f^\pm$, with $f^0,f^\pm$ being neutral and charged Standard Model fields, being the sum of the dark matter masses.

With $s_\delta=0$ and $\omega_{11}\sim 0$, the abundance is given as a function of the mass splitting. For each mass there exists a value $c > 1$, so that the coannihilations alone satisfy the abundance constraint. Fig.~\ref{SDM_z2_m1ceps} illustrates this critical value for $c$. For a parameter set $m_1,c$ above this critical line, a choice of the Higgs couplings and $s_\delta$ exists, so that the abundance constraint can be met. The red area is excluded by the abundance constraint in eq.~(\ref{eq-abundance}). For the $S_1$ mass in the range 1.33~TeV~$\leq m_1 \leq$~1.68~TeV, the coannihilations have the appropriate annihilation efficiency to match the abundance constraint, so that $c=1$ is allowed. 
\vspace{2mm}

\begin{figure}[h]
\begin{center}
\includegraphics[width=0.4\textwidth,angle=-90]{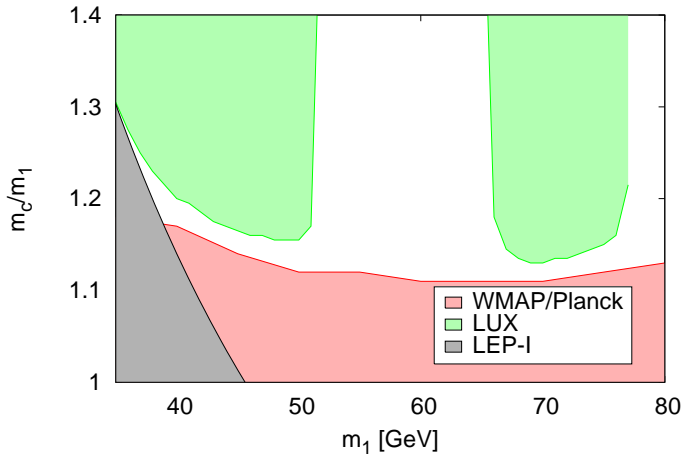}
\end{center}
\caption{Parameter space in the relaxed singlet model. The red area represents a \DM\, relic abundance being too small compared to the value reported by WMAP/Planck. The green area represents the combination of the abundance constraint from WMAP~\cite{Hinshaw:2012aka} and Planck~\cite{Ade:2013zuv} with the direct search constraints from LUX~\cite{Akerib:2013tjd}. The gray area denotes the decays of the $Z$ boson into $S^\pm$ violating LEP-I exclusion limits \cite{Beringer:1900zz}.}
\label{SDM_z2_relaxedsingleteps}
\end{figure}

\noindent
We define the {\it relaxed singlet scenario} with the following constraints: 
\begin{equation}
s_\delta < 0.3 \qquad {\rm and} \qquad c < c_{\rm max}\,.
\end{equation}
This definition includes the coannihilations of the $S^\pm$. In this way small values of $\omega_{11}$ are not necessarily excluded through the abundance constraint. We say that the abundance constraint on the Higgs coupling is relaxed. 

For $c$ close to border of the red area in Fig.~\ref{SDM_z2_m1ceps}, the abundance constraint on $\omega_{11}$ can become arbitrarily small. This allows for a spin-independent $S_1$ nucleon scattering cross section, sufficiently small not to violate the direct search exclusion limits from the LUX experiment for $m_1 \leq 53$ GeV and also for 66 GeV $\leq m_1 \leq$ 78 GeV.

The uncolored area in Fig.~\ref{SDM_z2_relaxedsingleteps}, framed by the exclusion limits from LUX, WMAP, Planck and LEP-I, can acommodate all constraints simultaneously. The LEP-I limits come from upper bounds on the $Z$ boson decay width as in eq.~(\ref{LEP-I}). The green area represents the value of $\sigma^{SI}$, as given by the abundance constraint on $\omega_{11}$, being excluded by the exclusion limits from the LUX experiment. The red area denotes coannihilations suppressing the abundance below the abundance constraint in eq.~(\ref{eq-abundance}). Thus, the lower bound on the mass $m_1$ is given by $\sim 35$ GeV in this scenario.

\vspace{2mm}

\begin{figure}[h]
\begin{center}
\includegraphics[width=0.4\textwidth,angle=-90]{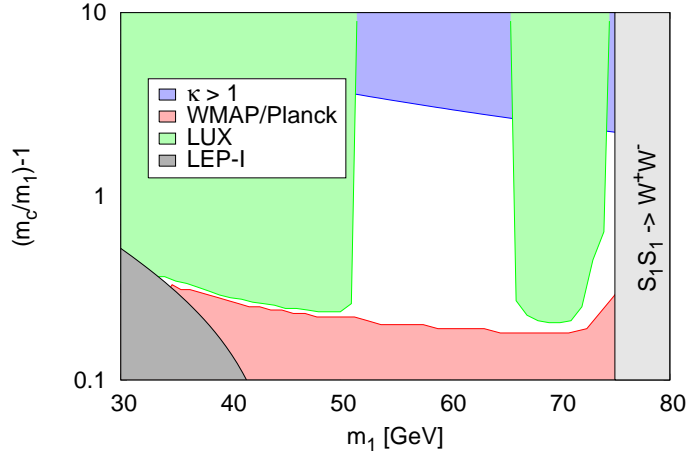}
\end{center}
\caption{Parameter space of the relaxed triplet scenario. The red area represents the WMAP/Planck exclusion limits, the green area denotes the combination of the abundance constraint and the direct search exclusion limits on the spin-independent $S_1$-nucleon scattering cross section ($\sigma^{SI})$ from LUX~\cite{Akerib:2013tjd}. The blue area denotes the upper bound from perturbativity and the gray area the suppression of the abundance by the coannihilations of the $S^\pm$, and the gray area the exclusion limits from LEP-I \cite{Beringer:1900zz}.}
\label{SDM_z2_tripletDSeps}
\end{figure}

\noindent
We define the {\it relaxed triplet scenario} by constraining
\begin{equation}
s_\delta > 0.75 \qquad {\rm and} \qquad 1.4 <  c < c_{\rm max}\,.
\end{equation}
This suppresses coannihilations and increases the sensitivity of the abundance to $\omega_{11}$. We notice, that the condition in eq.~(\ref{eq-m1bound}) limits $m_1$ from above with 238 GeV.

Fig.~\ref{SDM_z2_tripletDSeps} shows the areas of parameter space for which the abundance constraint in eq.~(\ref{eq-abundance}) can be met by a variation of $\omega_{11}$. For values of $c$ and $m_1$ as defined by the red area, the coannihilations are too efficient. The blue area denotes the perturbative constraint. For masses $m_1 \sim M_W$, the $W$ boson production threshold is reached. Due to virtual $W$ production, this threshold is given with 75 GeV. For $m_1$ at the threshold and beyond, the four-point interaction between the $S_1$ and the $W$ bosons dominates the total annihilation cross section efficiency, which is too large to match the abundance constraint for allowed masses. Thus the upper limit of $m_1$ in this scenario is given by 75 GeV. 

WMAP/Planck data constrains $\omega_{11}$ for $m_1,c$ in the allowed range, which fixes the value of $\sigma^{SI}$. The green area in Fig.~\ref{SDM_z2_tripletDSeps} shows the LUX exclusion limits on the parameter space. Altogether, the relaxed triplet scenario confines the mass $m_1$ to the range 51 GeV $\leq m_1 \leq$ 75 GeV. 

The absolute values of the mass splittings $\Delta_{ij}$ in this scenario can become very large. In particular $c = 3.2$ is possible, and with $m_2 = c/c_\delta$, $c_2$ is constrained by perturbativity. This leads to a decay spectrum with large mass splittings for $m_1 \simeq 51$ GeV and $s_\delta=0.75$
\begin{eqnarray}
51.1 \text{ GeV} &\leq \Delta_{21} & \leq 366 \text{ GeV}\,, \\
10.2 \text{ GeV} &\leq \Delta_{c1} &\leq 115 \text{ GeV}\,,\\
40.9 \text{ GeV} &\leq \Delta_{2c} & \leq 251 \text{ GeV}\,.
\end{eqnarray}

\section{Conclusions}
\label{sec_Conclusion}
In this paper we introduced the singlet-triplet $Z_2$ model. In this model an additional singlet and $SU(2)_L$ triplet field reside in the odd representation of a global $Z_2$ symmetry. Spontaneous symmetry breaking induces mass mixing among the neutral $Z_2$ odd fields, which renders the lightest one a suitable candidate for cold dark matter.

We investigated the constraints from WMAP and Planck data, the direct search experiment LUX, LEP-I and perturbativity on the model parameter space. The latter can be separated in three distinct scenarios. The first of these is the singlet scenario. It is comparable to the pure scalar singlet model, where the coupling between the Higgs boson and the \DM and the \DM mass parameter are the most relevant parameters. In this scenario the ranges of 51 GeV $\leq m_1 \leq 66$ GeV and 78 GeV $\leq m_1 \leq$ 3.4 TeV are allowed for the mass of the \DM candidate. The heavier neutral and the charged $Z_2$ odd particles are nearly mass degenerate, which results in detector signatures as difficult to detect at a collider as in the case of the pure triplet model~\cite{FileviezPerez:2008bj}. The mass splitting between $S^\pm$ and $S_1$, however, can be ${\cal O}(100$ GeV) for $m_1 \approx 1$ TeV, which allows for decays of the form $S^\pm \to S_1 f^\pm$, with $f^\pm$ being any Standard Model field with an even charge.

The second scenario is the triplet scenario, where coannihilations are relevant and the total annihilation cross section is dominated by the gauge couplings. This scenario is very close to the pure triplet model. The extra parameters extend the range for the \DM mass to 1.33 TeV $\leq m_1 \leq$ 6.65 TeV. Gamma ray data from the HESS collaboration can constrain this mass range further, however in a way that is highly dependent on the
model for the halo. The mass splitting between the neutral and charged components are bounded from above with 2.8 GeV for the lightest and charged fields, and to ${\cal O}(10$ GeV) for the other two combinations, respectively. 

The third scenario is an intermediate one, which includes two interesting sub scenarios: The relaxed singlet scenario demands a small mixing angle, but allows for the inclusion of coannihilations. This relaxes the abundance constraint on the $S_1$-Higgs coupling, which allows for a reduction of the lower bound on $m_1$ to $35$ GeV. In the relaxed triplet scenario, the neutral triplet component constitutes the dominant part of the \DM candidate, and coannihilations are suppressed. This leads to a range for the \DM mass between 51 and 75 GeV.

In the relaxed scenarios, some of the mass-splittings between the $S_1,S^\pm,S_2$ can become larger than $M_W$. This has consequences for collider searches, because a mass splitting above the $M_W$ threshold allows for the decays $S_2 \to S^\pm W^\mp$ and $S^\pm \to S_1 W^\pm$. These decays should lead to more detectable signatures at the LHC compared to the signals from the decay into  rather soft SM particles  in the triplet scenario.

\section*{Acknowledgements}
This work was supported by the Deutsche Forschungsgemeinschaft (DFG) through the Graduate School GRK 1102 "Physik an Hadron-Beschleunigern" and by the Bundesministerium f\"ur Bildung und Forschung within the F\"order-schwerpunkt
\textit{Elementary Particle Physics}. We thank Drs. Lorenzo Basso and Andrea Banfi for a careful reading of the manuscript, Prof. Dan Akerib and Dr. Blair Edwards from Yale for helpful assistance with the LUX data. Further we acknowledge the use of the Legacy Archive for Microwave Background Data Analysis (LAMBDA). Support for LAMBDA is provided by the NASA Office of Space Science.

\bibliographystyle{ieeetr}
\bibliography{library}

\begin{thebibliography}{10}

\bibitem{Zwicky:1933gu}
F.~Zwicky, ``{Spectral displacement of extra galactic nebulae},'' {\em
  Helv.Phys.Acta}, vol.~6, pp.~110--127, 1933.

\bibitem{Rubin:1970zza}
V.~C. Rubin and J.~Ford, W.~Kent, ``{Rotation of the Andromeda Nebula from a
  Spectroscopic Survey of Emission Regions},'' {\em Astrophys.J.}, vol.~159,
  pp.~379--403, 1970.

\bibitem{Silveira:1985rk}
V.~Silveira and A.~Zee, ``{Scalar Phantoms},'' {\em Phys.Lett.}, vol.~B161,
  p.~136, 1985.

\bibitem{McDonald:1993ex}
J.~McDonald, ``{Gauge singlet scalars as cold dark matter},'' {\em Phys.Rev.},
  vol.~D50, pp.~3637--3649, 1994.

\bibitem{Binoth:1994pv}
T.~Binoth and J.~van~der Bij, ``{Higgs signals modified by singlet scalars},''
  1994.

\bibitem{Binoth:1996au}
T.~Binoth and J.~van~der Bij, ``{Influence of strongly coupled, hidden scalars
  on Higgs signals},'' {\em Z.Phys.}, vol.~C75, pp.~17--25, 1997.

\bibitem{Burgess:2000yq}
C.~Burgess, M.~Pospelov, and T.~ter Veldhuis, ``{The Minimal model of
  nonbaryonic dark matter: A Singlet scalar},'' {\em Nucl.Phys.}, vol.~B619,
  pp.~709--728, 2001.

\bibitem{Boehm:2003hm}
C.~Boehm and P.~Fayet, ``{Scalar dark matter candidates},'' {\em Nucl.Phys.},
  vol.~B683, pp.~219--263, 2004.

\bibitem{Barbieri:2006dq}
R.~Barbieri, L.~J. Hall, and V.~S. Rychkov, ``{Improved naturalness with a
  heavy Higgs: An Alternative road to LHC physics},'' {\em Phys.Rev.},
  vol.~D74, p.~015007, 2006.

\bibitem{FileviezPerez:2008bj}
P.~Fileviez~Perez, H.~H. Patel, M.~Ramsey-Musolf, and K.~Wang, ``{Triplet
  Scalars and Dark Matter at the LHC},'' {\em Phys.Rev.}, vol.~D79, p.~055024,
  2009.

\bibitem{Cirelli:2005uq}
M.~Cirelli, N.~Fornengo, and A.~Strumia, ``{Minimal dark matter},'' {\em
  Nucl.Phys.}, vol.~B753, pp.~178--194, 2006.

\bibitem{Kadastik:2009dj}
M.~Kadastik, K.~Kannike, and M.~Raidal, ``{Matter parity as the origin of
  scalar Dark Matter},'' {\em Phys.Rev.}, vol.~D81, p.~015002, 2010.

\bibitem{vanderBij:2007fe}
J.~van~der Bij, ``{A Cosmotopological relation for a unified field theory},''
  {\em Phys.Rev.}, vol.~D76, p.~121702, 2007.

\bibitem{vanderBij:2010nu}
J.~van~der Bij, ``{Gravitational anomaly and fundamental forces},'' {\em Gen.
  Relativ. Gravit.}, 2010.

\bibitem{Aprile:2011hi}
E.~Aprile {\em et~al.}, ``{Dark Matter Results from 100 Live Days of XENON100
  Data},'' {\em Phys. Rev. Lett.}, vol.~107, p.~131302, 2011.

\bibitem{Fischer:2011zz}
O.~Fischer and J.~van~der Bij, ``{Multi-singlet and singlet-triplet scalar dark
  matter},'' {\em Mod.Phys.Lett.}, vol.~A26, pp.~2039--2049, 2011.

\bibitem{Belanger:2001fz}
G.~Belanger, F.~Boudjema, A.~Pukhov, and A.~Semenov, ``Micromegas,'' {\em
  Comput.Phys.Commun.}, vol.~149, pp.~103--120, 2002.

\bibitem{Belanger:2004yn}
G.~Belanger, F.~Boudjema, A.~Pukhov, and A.~Semenov, ``{micrOMEGAs: Version
  1.3},'' {\em Comput.Phys.Commun.}, vol.~174, pp.~577--604, 2006.

\bibitem{Beringer:1900zz}
J.~Beringer {\em et~al.}, ``{Review of Particle Physics (RPP)},'' {\em
  Phys.Rev.}, vol.~D86, p.~010001, 2012.

\bibitem{Hinshaw:2012aka}
G.~Hinshaw {\em et~al.}, ``{Nine-Year Wilkinson Microwave Anisotropy Probe
  (WMAP) Observations: Cosmological Parameter Results},'' {\em
  Astrophys.J.Suppl.}, vol.~208, p.~19, 2013.

\bibitem{Ade:2013zuv}
P.~Ade {\em et~al.}, ``{Planck 2013 results. XVI. Cosmological parameters},''
  2013.

\bibitem{Akerib:2013tjd}
D.~Akerib {\em et~al.}, ``{First results from the LUX dark matter experiment at
  the Sanford Underground Research Facility},'' 2013.

\bibitem{Alarcon:2012nr}
J.~Alarcon, L.~Geng, J.~Martin~Camalich, and J.~Oller, ``{On the strangeness
  content of the nucleon},'' 2012.

\bibitem{Abramowski:2011hc}
A.~Abramowski {\em et~al.}, ``{Search for a Dark Matter annihilation signal
  from the Galactic Center halo with H.E.S.S},'' {\em Phys.Rev.Lett.},
  vol.~106, p.~161301, 2011.

\bibitem{Fan:2013faa}
J.~Fan and M.~Reece, ``{In Wino Veritas? Indirect Searches Shed Light on
  Neutralino Dark Matter},'' {\em JHEP}, vol.~1310, p.~124, 2013.

\bibitem{Cohen:2013ama}
T.~Cohen, M.~Lisanti, A.~Pierce, and T.~R. Slatyer, ``{Wino Dark Matter Under
  Siege},'' {\em JCAP}, vol.~1310, p.~061, 2013.

\end{thebibliography}

\end{document}